# Effect of structure and composition on the electronic excitation induced amorphization of La$_2$Ti$_{2-x}$Zr$_x$O$_7$ ceramics

Michel Sassi[1*], Tiffany Kaspar[1], Kevin M. Rosso[1], and Steven R. Spurgeon[2]

[1]Physical and Computational Sciences Directorate, Pacific Northwest National Laboratory, Richland, Washington 99354, USA

[2]Energy and Environment Directorate, Pacific Northwest National Laboratory, Richland, Washington 99354, USA

*michel.sassi@pnnl.gov

**Abstract:** Understanding the response of ceramics operating in extreme environments is of interest for a variety of applications. *Ab initio* molecular dynamic simulations have been used to investigate the effect of structure and *B*-site (=Ti, Zr) cation composition of lanthanum-based oxides (La$_2$B$_2$O$_7$) on electronic-excitation-induced amorphization. We find that the amorphous transition in monoclinic layered perovskite La$_2$Ti$_2$O$_7$ occurs for a lower degree of electronic excitation than for cubic pyrochlore La$_2$Zr$_2$O$_7$. While in each case the formation of O$_2$-like molecules drives the structure to an amorphous state, an analysis of the polyhedral connection network reveals that the rotation of TiO$_6$ octahedra in the monoclinic phase can promote such molecule formation, while such octahedral rotation is not possible in the cubic phase. However, once the symmetry of the cubic structure is broken by substituting Ti for Zr, it becomes less resistant to amorphization. A compound made of 50% Ti and 50% Zr (La$_2$TiZrO$_7$) is found to be more resistant in the monoclinic than in the cubic phase, which may be related to the lower bandgap of the cubic phase. These results illustrate the complex interplay of structure and composition that give rise to the radiation resistance of these important functional materials.

**Introduction:**

Pyrochlores, with the general ideal formula $A_2B_2O_7$, represent an important class of complex oxide ceramics. With more than 500 different combinations of A- and B-site cations[1,2], pyrochlores exhibit a range of physical, chemical and electronic properties, making them useful for a variety of applications, including nuclear fuel and waste forms, solar energy conversion, electronics, and catalysis[1,2]. While the A-site cation is generally occupied by tri- and tetravalent actinides and lanthanides, the B-site is usually occupied by transition or post-transition metals. The stability and structural flexibility of pyrochlores is governed by the radius ratio between the A- and B-site cations[3]. $A_2B_2O_7$ oxides with a $r_A/r_B$ ratio ranging from 1.46 to 1.78 are isometric ($Fd\bar{3}m$, Z=8, a=0.9-1.2 nm) and adopt the pyrochlore structure, which is related to ideal fluorite ($AX_2$) but with ordered cation sites and anion vacancies. If the cation radius ratio is above 1.78, then a monoclinic layered perovskite structure ($P2_1$) is favored, while for a ratio below 1.46, a distorted fluorite structure ($Fd\bar{3}m$) is preferred.

Pyrochlores used in the nuclear fuel cycle will be exposed to extreme conditions potentially affecting their long-term performance and reliability. Pyrochlores are potential waste form candidates for the immobilization of plutonium and other minor actinides from spent nuclear fuel[4-7] because significant amounts of radioisotopes can be incorporated into the A-site of the pyrochlore structure. To better understand the structural changes that can occur under irradiation, pyrochlores have been the subject of many experiments that focused on characterizing their response to electron[8,9], ion[10,11], and photon (pulsed laser)[12,13] irradiation in terms of radiation-induced defect generation, damage processes, microstructural evolution, and phase transformations. Many studies[14-24] were specifically designed to investigate the effect of α-decay in various pyrochlore compositions using keV-GeV ion-beam irradiation; in this range, ion stopping is mediated predominantly by interactions with the target electrons, inducing excitations and ionizations along the ion path. These experiments have highlighted the importance of structure, bond-type, and electronic configuration on their radiation stability, particularly for the titanate ($A_2Ti_2O_7$) and zirconate ($A_2Zr_2O_7$) pyrochlores[15,21-24].

While pyrochlore structures display a wide range of behaviors in response to ion-beam irradiation as a function of composition, it is generally found that titanate pyrochlores, such as

Gd$_2$Ti$_2$O$_7$, are less resistant to electronic excitations compared to their zirconate counterparts, such as Gd$_2$Zr$_2$O$_7$. Analysis of the thermal spike following 119 MeV U irradiation[15] shows that Gd$_2$Ti$_2$O$_7$ is readily amorphized, while Gd$_2$Zr$_2$O$_7$ is transformed into a radiation-resistant anion-deficient fluorite structure. For other zirconate pyrochlores ($A_2$Zr$_2$O$_7$), a more complex response was observed, in which both a pyrochlore-fluorite phase transformation and amorphization occur. *B*-site mixed Gd$_2$Ti$_{2-x}$Zr$_x$O$_7$ compounds showed a systematic increase in the resistance to ion-beam induced amorphization with increasing Zr content[15,16,23]. This result highlights the important effect of the *B*-site cation on the response of pyrochlore materials to electronic excitations. Although zirconate pyrochlores are generally considered to be radiation-resistant, a systematic ion-beam study of $A_2$Zr$_2$O$_7$ pyrochlores shows that the susceptibility to amorphization by electronic excitation increases with increasing cation radius ratio $r_A/r_B$[15].

While theoretical studies[25-29] investigating the effect of atomic collision processes, defect generation and phase transformation have been performed to understand the mechanisms leading to amorphization, the role of electronic excitation in amorphization has been investigated less. The simulation of electronic excitation effects is a challenging task, in which we must account for the dynamics of the electron-hole recombination[30]. In order to mimic the effects of electronic excitation and ionization, a simplified methodology, consisting of removing several electrons from the higher valence band states, has been proposed and used in previous studies to investigate electronic excitations induced amorphization of Ge-Sb-Te alloys[30] and titanate pyrochlores[31]. It is worth noting that the effect of electronic excitation described by this method, and the resulting mechanisms leading to amorphization, are different from the intense electronic excitation induced by swift heavy ion irradiation, where local melting from the thermal spike leads to a quenched melt structure[31,32]. Nevertheless, these simulations of electronic excitation provide guidance for intense electron and pulsed laser irradiation studies, and help interpret trends in the susceptibility to electronic excitation-induced amorphization.

In this study, we report *ab initio* molecular dynamics simulations that focus on determining how the transition temperature from crystalline to amorphous phase is affected by the structure, composition and electronic excitation concentration in lanthanum-based compounds. In that regard, La$_2$Ti$_2$O$_7$ and La$_2$Zr$_2$O$_7$ are ideal materials to investigate since they respectively occur in monoclinic layered perovskite and cubic pyrochlore structures. Therefore, mixing the *B*-site cation modifies not only the composition of the material, but also its structure. In this study we also consider intermediate monoclinic $La_2Ti_{2-x}Zr_xO_7$ and cubic $La_2Zr_{2-x}Ti_xO_7$ compounds to follow the evolution of the amorphization temperature as function of the *B*-site composition. We find that differences in the octahedral connectivity of the cubic and monoclinic phases significantly influence the potential for defect generation and relative phase stability, with important implications for the design of nuclear energy systems.

**Computational Details:**

In order to investigate the effect of structure and composition on electronic excitation induced amorphization, several compounds with variable *B*-site cation composition have been created by using the special quasirandom structure (SQS)[33] generation code available in the ATAT toolkit[34]. Two series of SQS intermediate structures were created. One series uses the monoclinic layered perovskite $La_2Ti_2O_7$ (LTO) crystal as its starting point and substitutes Zr for Ti, giving rise to $La_2Ti_{2-x}Zr_xO_7$ compounds (labeled Z-LTO) with a *monoclinic* lattice. The other series uses the cubic pyrochlore $La_2Zr_2O_7$ (LZO) crystal as its starting point and substitutes Ti for Zr, leading to $La_2Zr_{2-x}Ti_xO_7$ compounds (labeled T-LZO) with a *cubic* lattice. For consistency with the end-member LTO and LZO crystals, only SQS structures containing a total of 88 atoms were generated, such that all of the structures possess 16 La sites, 16 *B*-sites occupied by either pure Ti, pure Zr, or a mix of both species, and 56 O atoms.

Density functional theory (DFT) calculations were performed on all the structures using the VASP package[35]. The exchange-correlation functional used the generalized gradient approximation (GGA) as parametrized by Perdew and Wang[36]. All of the calculations accounted for spin-polarization, used the Vosko-Wilk-Nusair local density approximation scheme[37], and a cutoff energy for the projector augmented wave[38] pseudo-potential of 400 eV. Prior to performing *ab initio* molecular dynamics simulations, the atomic coordinates and lattice parameters of each investigated structure were fully relaxed using a convergence criterion of $10^{-5}$ eV/cell for the total energy and $10^{-4}$ eV/Å for the force components. A 4×4×2 and 4×4×4 Monkhorst-Pack[39] *k*-point sampling of the Brillouin zone was used for the monoclinic and cubic structures respectively.

To mimic the effect of electronic excitation, several electrons have been removed from the higher valence band states and a jellium background has been used to compensate for the loss of charge. A similar procedure has been used previously to investigate the role of electronic excitation in the amorphization of Ge-Sb-Te alloys[30] and titanate pyrochlores[31]. The electronic excitation concentration has been defined as the percentage ratio of removed electrons with respect to the total number of electrons in the system. For the monoclinic LTO and cubic LZO crystals, the total number of electrons for a cell containing 88 atoms was 1728 and 2000, respectively, and the number of removed electrons varied from 7 to 35 for LTO and 8 to 40 for LZO, such that the excitation concentration investigated in the structures varied from 0.4% to 2% by steps of 0.4%. The electronic excitation concentration in the intermediate SQS structures followed a similar calculation. To determine the temperature at which the crystalline to amorphous transition occurs for each structure and electronic excitation conditions, *ab initio* molecular dynamic simulations have been performed at the Γ-point of the Brillouin zone in the NVT ensemble with a Nosé-Hoover thermostat for temperatures ranging from 200 K to 1000 K by step of 100 K. For each case, a three-step procedure was followed for a total simulation time of 27 ps with a time step of 2 fs. First, 6 ps were used to thermalize the crystal at the given temperature. Second, 15 ps were given to allow structural modifications associated with electronic excitation. This time lapse was tested in each structure and proven long enough to allow the system to reach an equilibrium state. Third, 6 ps were used to allow electron-hole recombination.

**Results and Discussion:**

The monoclinic LTO structure is a layered perovskite crystal containing $TiO_6$ octahedra linearly connected, as shown in Fig. 1(a). In contrast, the cubic pyrochlore LZO crystal contains $ZrO_6$ octahedra inter-connected by corners in a zigzag fashion. A comparison of the two crystals in their respective ground state structure indicates that the monoclinic LTO is 0.325 eV/atom less energetically favorable than the cubic LZO, as shown in Fig. 1(b). In the case where a monoclinic perovskite structure is made of 100% Zr atoms (*i.e.*, monoclinic LZO) it would be 0.039 eV/atom less favorable than its ground state cubic LZO structure. Similarly, if a cubic pyrochlore crystal was made of 100% Ti atoms (*i.e.*, cubic LTO), its energy would be 0.055 eV/atom less favorable than that of the monoclinic LTO crystal. The general trend shown in Fig. 1(b) is that the incorporation of Zr in both crystal structures leads to more energetically favorable phases. For both the monoclinic and cubic crystals, SQS structures have been generated to find compositionally intermediate structures with the most appropriate [Ti, Zr] distributions that mimic, for a small number of atoms, the first few, physically most relevant radial correlation functions of a perfectly random structure. Using this method enables us to investigate the effect of progressive [Ti, Zr] mixing on electronic excitation induced amorphization. Among the SQS structures generated for 25%, 50% and 75% of Zr (calculated from the *B*-site only), only those with the lowest energy have been reported in Fig. 1(b). The energy diagram shows that the phase transition from monoclinic perovskite to cubic pyrochlore should occur for 53% of Zr in the monoclinic LTO structure. It is worth noting that the structures at 50% [Ti, Zr] mixing have very similar energies but different crystal structures. In the following, we have focused our investigations on the most energetically stable structures at any given composition, mainly by considering only a monoclinic layered perovskite phase for compounds with a Zr fraction ≤50% and a cubic pyrochlore phase for compounds with a Zr fraction ≥50%. At a Zr fraction of 50% both phases have been investigated.

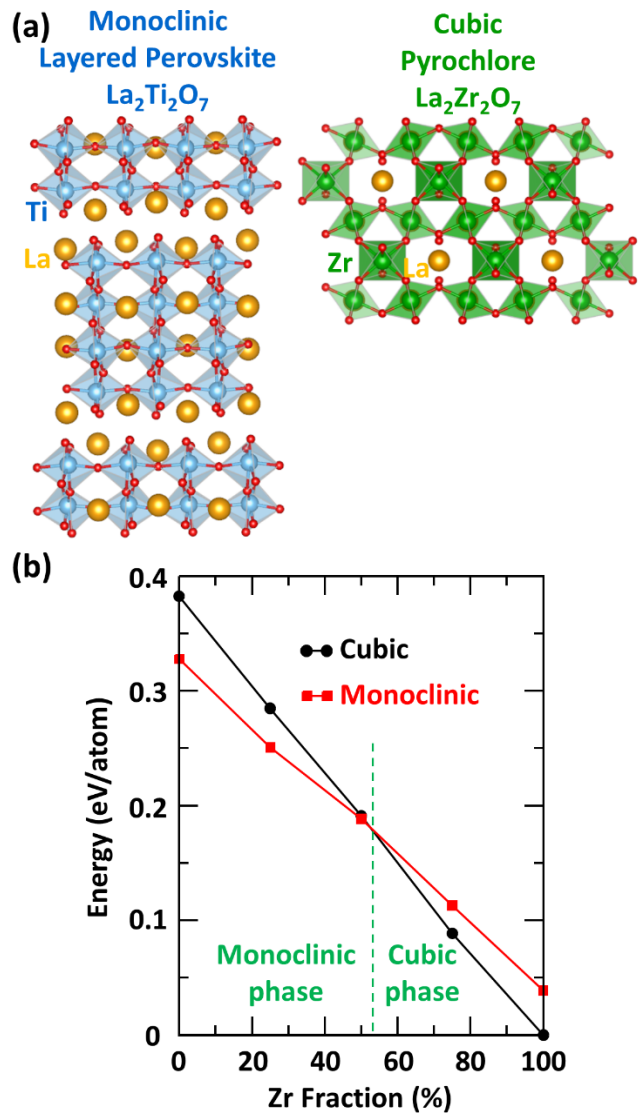

Figure 1: (a) Representation of the monoclinic layered perovskite LTO and cubic pyrochlore LZO structures. (b) Evolution of the crystal energy as function of Zr fraction in the monoclinic and cubic lattices.

To determine how the crystal structure, composition and electronic excitation concentration affect the transition temperature from crystalline to amorphous phase, the radial distribution function (RDF) has been calculated at the end of the simulation time at each temperature and for each structure investigated. A material is considered amorphous when the short-range order in its RDF is maintained but the long-range order is lost. Fig. 2(a) shows the effect of various electronic excitation concentrations on the RDF of cubic LZO at 300 K. The features of the RDF remain similar until an excitation concentration of 1.2%, at which point bonds between oxygen atoms start to form. The appearance of $O_2$-like molecules is associated with a modification of the intermediate and long-range ordering, suggesting the beginning of amorphization. At higher electronic excitation concentrations, the degree of amorphization increases with the number of O—O bonds formed. It is worth noting that a similar behavior has been obtained by Xiao *et al.*[31] during the simulation of the effect of electronic excitation in several titanate pyrochlores. Under the influence of electronic excitation, the formation of $O_2$-like molecules in these materials is due to the electrons removed from the system, which mainly originate from oxygen 2*p* valence orbitals located near the Fermi level. The anion disorder resulting from the formation of $O_2$ molecules further induces cation displacements, together driving the structure toward an amorphous phase. Fig. 2(b) shows the evolution of the RDF features at different temperatures for an excitation concentration fixed at 0.8%. The RDFs show that the formation of O—O bonds starts at 600 K, which coincides with the beginning of amorphization.

Performing similar calculations for other excitation concentrations and [Zr, Ti] mixing allows us to investigate how the *B*-site composition affect the transition temperature from crystalline to amorphous phase, and its interplay with electronic excitation concentration in the cubic pyrochlore lattice. For each composition investigated, the general trend, shown in Fig. 2(c), is a lowering of the transition temperature as the excitation concentration increases such that each cubic compound amorphizes at room temperature for 1.2% of excitation concentration. While cubic LZO is found to be the most resistant composition to electronic excitation, a compound made of 50% Zr (i.e. $La_2ZrTiO_7$) is the easiest to amorphize as suggested by the value of its transition temperature, which is always the lowest across the range of electronic excitation concentration investigated. Interestingly, a similar trend with respect to *B*-site composition has been observed experimentally for the $Gd_2(Zr_xTi_{1-x})_2O_7$ pyrochlore family, with the pure $Gd_2Zr_2O_7$ phase being more resistant to electronic excitation than a phase made of 50% Zr fraction (i.e. $Gd_2ZrTiO_7$)[16]. The evolution of the transition temperature as function of Zr fraction for 0.4%, 0.8% and 1.2% of excitation concentration is shown in Fig. 2(d). While for 0.4% of excitation concentration, the transition temperature is highly dependent on the *B*-site cation composition, for 1.2% of excitation concentration the amorphization temperature starts to be composition-independent.

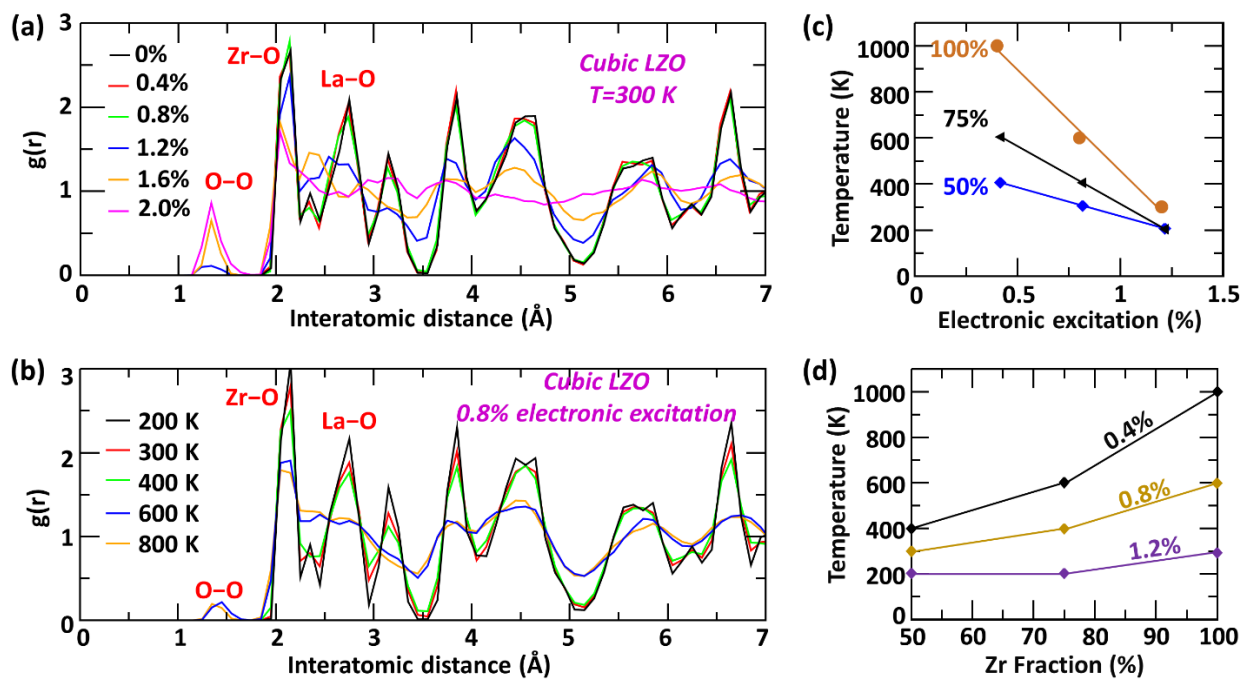

Figure 2: Evolution of the radial distribution function (g(r)) of cubic LZO as function of (a) the electronic excitation concentration at 300 K and (b) the temperature for a fixed 0.8% of electronic excitation concentration. (c) Evolution of crystalline-to-amorphous transition temperature in cubic T-LZO compounds as function of the excitation concentration for 50%, 75% and 100% Zr fraction. (d) Evolution of crystalline-to-amorphous transition temperature as function of Zr fraction for 0.4%, 0.8% and 1.2% excitation concentration.

The response of the monoclinic layered perovskite compounds to electronic excitation and the variations of the amorphization temperature with respect to *B*-site composition are very different than those of cubic pyrochlore compounds. The evolution of the RDF features for monoclinic LTO at 300 K for three excitation concentrations is shown in Fig. 3(a). As for the cubic compounds, the formation of O—O bonds is maintained, but at an excitation concentration of 0.8%, which is lower than the 1.2% determined for the cubic lattice. Although 0.4% excitation concentration is not enough to have created O—O bonds in the crystal, the RDF features have been slightly affected compared to monoclinic LTO without electronic excitation (0%). At 0.4% electronic excitation, no Ti—O or La—O bonds have been broken, but their distances have varied by 0.1-0.2 Å on average, resulting in a slightly broadened and less intense RDF short-range peak for Ti—O. As shown in Fig. 3(b), the transition temperature tends to decrease in each monoclinic Z-LTO compound as the excitation concentration increases. Interestingly, the monoclinic layered perovskite lattice is most resistant to electronic excitation when the chemical mixing in the *B*-site is the highest, which is obtained for 50% of Zr fraction. This is in contrast to the cubic pyrochlore lattice, for which the most vulnerable compound is obtained for a 50% [Zr, Ti] mixing. For 1.2% of excitation concentration and above, the transition temperature is composition independent and all the monoclinic compounds are readily amorphized below room temperature.

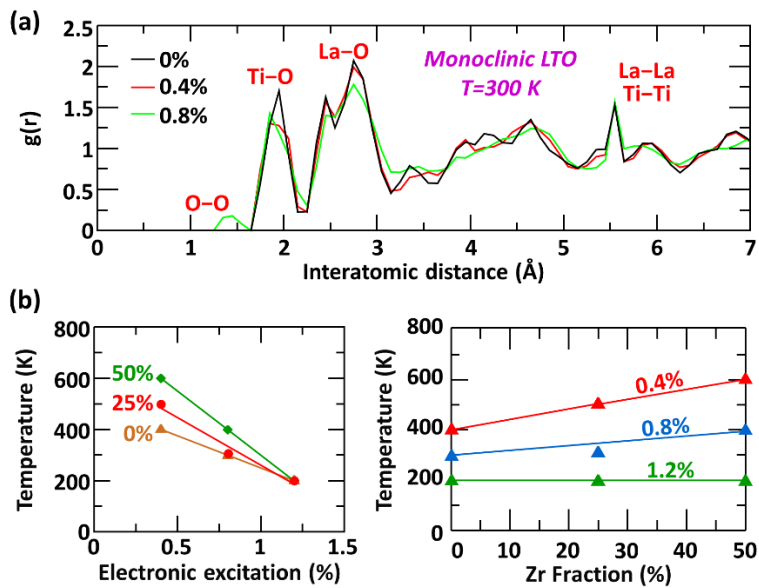

Figure 3: (a) Effect of electronic excitation concentration on the radial distribution function (g(r)) of the monoclinic LTO crystal at 300 K. (b) Evolution of transition temperature as function of excitation concentration and composition for monoclinic Z-LTO compounds.

A direct comparison of the overall interplay between structure, composition and electronic excitation concentration on transition temperature is shown in Fig. 4. For both 0.4% and 0.8% excitation concentration, the monoclinic compound made of 50% [Zr, Ti] mixing is found to be more resistant than cubic compound of the same composition. Fig. 4 also shows that the variations of the transition temperature with Zr fraction are larger for the cubic phase than for the monoclinic phase. This behavior suggests that the introduction of Ti in the cubic LZO phase has a greater impact on its resistance than the introduction of Zr in the monoclinic LTO phase. Breaking the high symmetry of the cubic phase by mixing the *B*-site composition has a more destabilizing effect than having a mixed *B*-site composition in the already low symmetry monoclinic phase. Interestingly, since the transition from monoclinic to cubic phase occurs at 53% Zr fraction, Fig. 4 shows that the 50% [Zr, Ti] composition of the cubic lattice, which is the most vulnerable to electronic excitation, should not be involved. This has been symbolized by the black dashed lines in Fig. 4. Therefore, starting from a monoclinic layered perovskite LTO structure, the gradual substitution of Ti by Zr would yield structures that are increasingly more resistant to electronic excitations, with the cubic pyrochlore LZO being the most resistant to electronic excitation. This overall trend is in agreement with several experimental observations[15,16,24], which found that the susceptibility to amorphization by electronic excitation increases with cation radius ratio $r_A/r_B$ and Zr content. In the case of cubic LZO and monoclinic LTO, the $r_A/r_B$ ratio is 1.61 and 1.92 respectively, making the LZO pyrochlore phase more resistant to electronic excitation. It is worth noting that the increase in electronic excitation resistance shown in Fig. 4 is not monotonic; a monoclinic compound with 50% Zr is predicted to have a similar transition temperature to a cubic compound with 75% Zr. While only the monoclinic and cubic phases were considered to model the phase transition between LTO and LZO, the potential for other phases to be involved in the transition region could lead to some uncertainty in the amorphization behavior in this regime. We also note that the trend in the resistance to electronic excitation follows the variations of the band gap ($E_g$) of the materials, which increases with Zr fraction. In particular, the band gap of the monoclinic phase with 50% Zr is slightly larger than the band gap of the cubic phase with 50% Zr, such that in order, $E_{g\,\text{Monoclinic}}^{\text{LTO}} < E_{g\,\text{Cubic}}^{\text{T-LZO}} < E_{g\,\text{Monoclinic}}^{\text{Z-LTO}} < E_{g\,\text{Cubic}}^{\text{LZO}}$. This correlation supports the idea that the radiation response cannot be solely described in terms of cation radius ratio and that other factors, such as defect formation energy, bond type (i.e. covalency and ionicity), electronic structure of the compounds, and energy of order-disorder transition, also play an important role in the amorphization resistance of ceramics to electronic excitation[40].

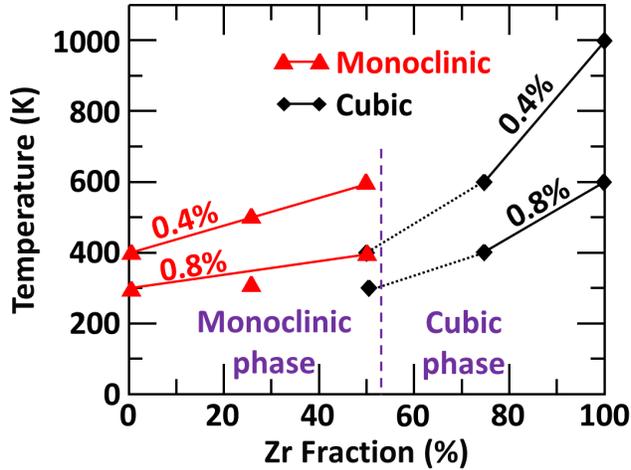

Figure 4: Effect of structure, composition and excitation concentration on the crystalline-to-amorphous transition temperature from the monoclinic perovskite to cubic pyrochlore structure.

A comparison of our results with similar theoretical calculations performed for titanate pyrochlores by Xiao *et al.*[31] complements and deepens our understanding of the effect of structure and composition on electronic excitation induced amorphization in ceramics. Especially, Fig. 5(a) shows that the evolution of transition temperature with excitation concentration for the monoclinic layered perovskite LTO has a similar slope compared to the other titanate pyrochlores, while cubic LZO pyrochlore exhibits a steeper slope. This suggests that a chemical change of the *B*-site cation more strongly affects the slope of the transition temperature as a function of excitation concentration, compared to a chemical change of the *A*-site cation. Fig. 5(a) also shows that cubic pyrochlores are more resistant to electronic excitation than the monoclinic perovskite LTO structure, as evidenced by their higher transition temperatures. In the specific case of a comparison between $La_2Ti_2O_7$ and $Gd_2Ti_2O_7$, experiments have shown that the perovskite structure amorphizes more rapidly than the pyrochlore structure[24]. This is in agreement with LTO having a lower crystalline-to-amorphous transition temperature compared to $Gd_2Ti_2O_7$.

An analysis of the structural modifications obtained at the end of the simulations for cubic LZO and monoclinic LTO allows us to identify the main structural features conferring resistance to electronic excitation. Fig. 5(b) shows an example of structural modifications obtained at 600 K and 0.8% excitation concentration for both cubic LZO and monoclinic LTO crystals. For the cubic

LZO crystal, the zigzag corner sharing ZrO$_6$ octahedral arrangement is mostly preserved, even if some O$_2$-like molecules have been formed in the structure. However, in the case of the monoclinic LTO crystal, the formation of O$_2$-like molecules more strongly disturbs the octahedral connectivity. Due to the breaking of Ti—O bonds, new La-La layers can form in addition to those initially present in the structure, and the formation of TiO$_5$ square based pyramids is possible. The structural origin of these modifications is the possibility for TiO$_6$ octahedra to rotate along the axis by which they are linearly inter-connected. Octahedral rotations allow the monoclinic LTO structure to accommodate electronic excitation with more flexibility, but they also facilitate the formation of O$_2$-like molecules, making the amorphization easier; ultimately, this results in the structure being less resistant to excitation. In the cubic pyrochlore structures, such octahedral rotation is prevented by the zigzag connectivity network between ZrO$_6$ octahedra.

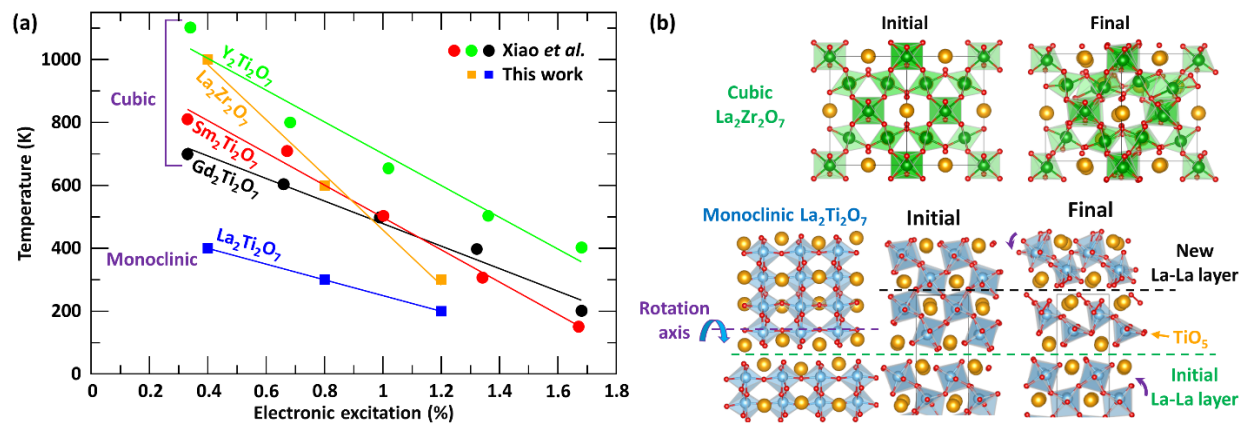

Figure 5: (a) Crystalline-to-amorphous transition temperature as function of electronic excitation for the lanthanum-based pyrochlores and comparison with the results from Ref[31]. (b) Example of structural deformations obtained for cubic LZO and monoclinic LTO crystals at a temperature of 600 K and an excitation concentration of 0.8%.

**Conclusions:**

*Ab initio* molecular dynamics simulations investigating the interplay between electronic excitation, structure and composition on the amorphization temperature of lanthanum-based ceramics have been performed. The effect of *B*-site composition is that the monoclinic layered perovskite phase is more resistant to excitation for 50% [Zr, Ti] mixing, while the cubic pyrochlore phase is more vulnerable for a high [Zr, Ti] mixing. A direct structural comparison between the monoclinic and cubic phases indicates that at 50% [Zr, Ti] mixing, the monoclinic perovskite structure is most resistant; however, the undoped cubic LZO crystal is found to be more resistant to excitation than the undoped monoclinic LTO crystal. A structural analysis suggests that the structural feature conferring excitation resistance in cubic pyrochlore crystals is the octahedral connectivity network with a zigzag pattern. In contrast, the rotation of linearly connected octahedra in the monoclinic LTO phase facilitates the formation of $O_2$-like molecules in the structure. Our results illustrate how crystal structure and octahedral distortions mediate the amorphization process, suggesting ways to improve the radiation response of important functional ceramics and waste forms.

**Acknowledgement:**

This work was supported by the Nuclear Process Science Initiative (NPSI) at the Pacific Northwest National Laboratory (PNNL). Computational resources were provided by PNNL Institutional Computing (PIC). This work was performed at Pacific Northwest National Laboratory which is operated by the Battelle Memorial Institute for the U. S. Department of Energy under Contract No. DEAC05-76RL0-1830.

**Author contributions:**

M.S. designed and conduced the calculations. All the authors discussed the results, wrote and reviewed the manuscript.

**Additional information:**

The authors declare no competing financial interests.